\documentclass[prl,twocolumn,superscriptaddress,citeautoscript,longbibliography,amsmath,amssymb,nobibnotes]{revtex4-2}

\usepackage{graphicx}
\usepackage{xcolor}
\usepackage{hyperref}

\graphicspath{{./}}

\newcommand{\EF}{$E_\mathrm{F}$}
\newcommand{\Gb}{$\bar{\Gamma}$}
\newcommand{\Xb}{$\bar{X}$}
\newcommand{\Mb}{$\bar{M}$}
\newcommand{\invA}{\AA$^{-1}$}

\begin{document}
% ============================================================

\title{Angle-Resolved Photoemission as a Sensitive Probe of Surface Quality:
       Sputtered versus Annealed Cu(001)}

% ---- co-first authors ----
\author{Hyeonhee Na}
\thanks{These authors contributed equally to this work.}
\affiliation{Electronic Structure Control (ESC) Lab, Department of Physics, Sungkyunkwan University, Suwon 16419, Republic of Korea}

\author{Jeonghwan Oh}
\thanks{These authors contributed equally to this work.}
\affiliation{Electronic Structure Control (ESC) Lab, Department of Physics, Sungkyunkwan University, Suwon 16419, Republic of Korea}

\author{Jinju Lee}
\thanks{These authors contributed equally to this work.}
\affiliation{Electronic Structure Control (ESC) Lab, Department of Physics, Sungkyunkwan University, Suwon 16419, Republic of Korea}

\author{Hanjun Sim}
\thanks{These authors contributed equally to this work.}
\affiliation{Electronic Structure Control (ESC) Lab, Department of Physics, Sungkyunkwan University, Suwon 16419, Republic of Korea}

% ---- co-authors ----
\author{Myeongun Go}
\affiliation{Electronic Structure Control (ESC) Lab, Department of Physics, Sungkyunkwan University, Suwon 16419, Republic of Korea}

\author{Jimin Sohn}
\affiliation{Electronic Structure Control (ESC) Lab, Department of Physics, Sungkyunkwan University, Suwon 16419, Republic of Korea}

\author{Jeonggu Kang}
\affiliation{Electronic Structure Control (ESC) Lab, Department of Physics, Sungkyunkwan University, Suwon 16419, Republic of Korea}

\author{Byungmin Sohn}
\email{bsohn@skku.edu}
\affiliation{Electronic Structure Control (ESC) Lab, Department of Physics, Sungkyunkwan University, Suwon 16419, Republic of Korea}

\date{\today}

% ------------------------------------------------------------
\begin{abstract}
Angle-resolved photoemission spectroscopy (ARPES) requires a well-ordered surface, whose quality is routinely judged from low-energy electron diffraction (LEED) and x-ray photoelectron spectroscopy (XPS). We compare LEED, XPS, and ARPES of a Cu(001) single crystal measured \textit{in situ} at four stages of surface preparation: as-prepared, after aging for 24~h at 6~K, after Ar$^+$ sputtering, and after annealing. The sputtered and annealed surfaces show nearly identical (1$\times$1) LEED patterns and Cu 2$p$ core levels, but the sputtered surface yields a momentum-independent ARPES spectrum, and the band dispersion recovers after annealing. We attribute this contrast to how each probe responds to sputtering-induced defects: LEED reflects the average lattice, which remains periodic, whereas ARPES reflects the electronic states, whose crystal momentum is no longer well defined. Our results indicate that a sharp LEED pattern and clean core-level spectra do not guarantee a surface suitable for ARPES.

\end{abstract}

\maketitle

% ============================================================
\section{Introduction}
% ============================================================

Angle-resolved photoemission spectroscopy (ARPES) measures the energy--momentum dispersion of electrons in a solid and has become the standard tool for determining the band structure of crystalline materials~\cite{Hufner2003,Damascelli2003,Sobota2021,Sohn2024,Sohn2025,Kim2026,Lee2026}. Because the photoelectrons escape predominantly from the near-surface region~\cite{SeahDench1979} and their momentum parallel to the surface is conserved during photoemission, ARPES requires a clean and well-ordered surface. Consequently, a substantial part of many ARPES experiments is devoted to preparing such a surface. For metal single crystals such as Au, Ag, and Cu, in which a clean surface cannot be obtained by standard cleaving techniques, the surface is typically prepared by repeated cycles of ion sputtering and annealing~\cite{Ritter1996}. Its quality is then monitored by low-energy electron diffraction (LEED) and x-ray photoelectron spectroscopy (XPS): sharp diffraction spots are taken as evidence of long-range order, and the absence of contaminant core levels as evidence of cleanliness~\cite{Woodruff1994}. LEED and XPS provide complementary information on surface structure and chemical composition and have been combined with ARPES in surface-characterization studies~\cite{Sohn2021,SohnKim2022,TlGe111,SiC6H,AgMo110,OMo110}.

It is less commonly appreciated, however, that a surface can satisfy both criteria and still fail to show band dispersion in ARPES. The three techniques probe different aspects of the surface: LEED is sensitive to the average periodic arrangement of the surface atoms~\cite{Henzler1982}, XPS core levels to the chemical composition~\cite{Chastain1992}, and ARPES to whether the electronic states in the topmost layers retain a well-defined in-plane crystal momentum~\cite{Sobota2021}. For example, on Cu(111), sputter-induced disorder broadens the surface-state features in photoemission, along with the LEED spots~\cite{Theilmann1997}. Nevertheless, a side-by-side comparison of LEED, XPS, and ARPES on the same surface before and after sputtering is rarely reported. Such a comparison is necessary for judging the extent to which the usual LEED and XPS criteria are reliable.

Here, we present such a comparison for Cu(001) [Figs.~\ref{fig:fig1}(a)--\ref{fig:fig1}(c)], whose bulk and surface electronic structures are well established~\cite{Courths1984,Knapp1979} and which therefore serves as a reference system. Using an \textit{in situ} cluster in which ARPES, XPS, LEED, and surface preparation share a single ultrahigh-vacuum (UHV) environment [Figs.~\ref{fig:fig1}(d) and \ref{fig:fig1}(e)], we measure the same crystal at four stages of preparation: as-prepared, after aging for 24~h at 6~K, after Ar$^+$ sputtering, and after annealing. We find that the sputtered and annealed surfaces are nearly indistinguishable in LEED and in the Cu 2$p$ core levels, whereas the ARPES spectrum of the sputtered surface loses its momentum dependence entirely and recovers after annealing. Our results show that ARPES can be more sensitive to the surface order than LEED or XPS, and that the ARPES spectrum itself provides the most direct test of whether a surface is suitable for band-structure measurements.

% ============================================================
\section{Methods}
% ============================================================

% ============================================================
%  Fig1
% ============================================================
\begin{figure*}[t]
  \includegraphics[width=\textwidth]{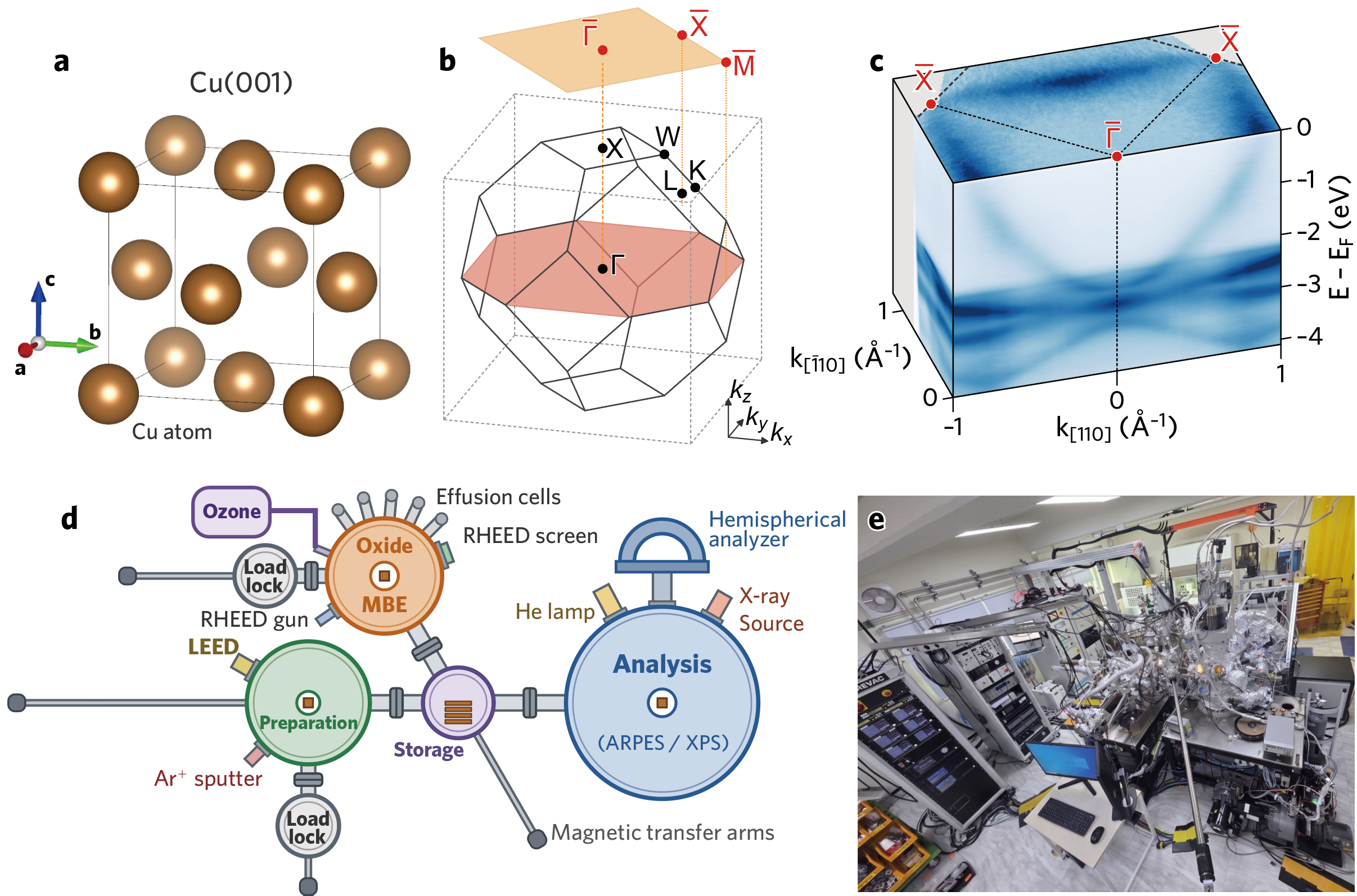}
\caption{\textbf{Crystal structure, Brillouin zone, angle-resolved photoemission spectroscopy (ARPES) data, and experimental setup.}
  (a) Face-centered cubic (fcc) unit cell of Cu ($a = 3.615$~\AA).
  (b) Bulk Brillouin zone and its projection onto the (001) surface Brillouin zone, with the high-symmetry points \Gb, \Xb, and \Mb.
  (c) Three-dimensional ARPES data volume ($E$, $k_{[110]}$, $k_{[\bar{1}10]}$) of the annealed Cu(001) surface, measured with He-I$\alpha$ photons ($h\nu = 21.2$~eV) at 6~K. The top face is the constant-energy map at \EF; the dashed line marks the surface Brillouin zone boundary, and \Gb\ and \Xb\ are indicated. The side faces are $E$--$k$ cuts through \Gb\ along the two in-plane axes.
  (d) Schematic top view and (e) photograph of the \textit{in situ} ARPES cluster. The preparation chamber [sputtering, annealing, and low-energy electron diffraction (LEED)], the oxide molecular beam epitaxy chamber, and the analysis chamber [ARPES and X-ray photoelectron spectroscopy (XPS)] are connected under ultrahigh vacuum (UHV), so that the sample is transferred between them without exposure to air.}
  \label{fig:fig1}
\end{figure*}

All measurements are performed in a single UHV cluster in which the analysis chamber for ARPES and XPS, the preparation chamber for ion sputtering, annealing, and LEED, an oxide molecular beam epitaxy (MBE) chamber, and the load locks are connected through a UHV transfer line [Figs.~\ref{fig:fig1}(d) and \ref{fig:fig1}(e)]. The sample is therefore transferred between surface treatment and measurement without exposure to ambient conditions, so that the LEED, XPS, and ARPES data presented below are obtained from the same surface at each stage of preparation. The base pressures of the analysis and preparation chambers are $2 \times 10^{-11}$ and $8 \times 10^{-10}$~Torr, respectively. 
% ============================================================
%  Fig2
% ============================================================
\begin{figure*}[t]
  \includegraphics[width=\textwidth]{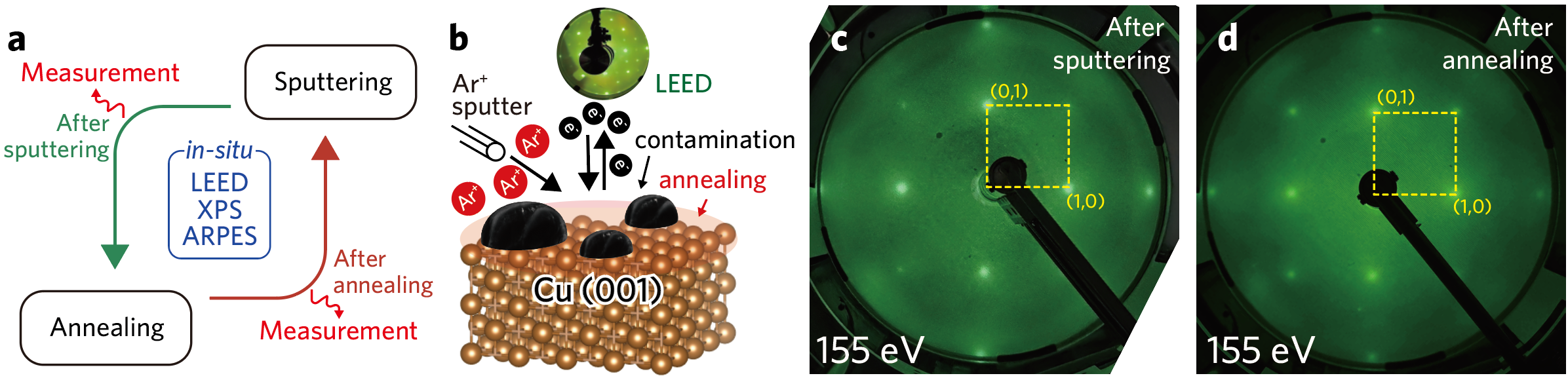}
  \caption{\textbf{Surface preparation cycle and LEED.}
  (a) Measurement sequence. The Cu(001) surface is repeatedly sputtered and annealed, and LEED, XPS, and ARPES are measured \textit{in situ} after sputtering (green) and after annealing (red).
  (b) Schematic of the surface treatment: Ar$^{+}$ sputtering removes surface contaminants but leaves the surface disordered, and subsequent annealing restores the surface order.
  (c),(d) LEED patterns taken at a beam energy of 155~eV after sputtering (c) and after annealing (d). The red dotted squares mark the $(1\times1)$ surface unit cell in reciprocal space, spanned by the $(1,0)$ and $(0,1)$ spots. Sharp $(1\times1)$ spots are observed in both cases.
  }
  \label{fig:fig2}
\end{figure*}
% ============================================================

The measurement sequence is schematically shown in Fig.~\ref{fig:fig2}(a). A Cu(001) single crystal (MTI Corporation, 99.99\%) is used throughout this work. The surface is prepared by repeated cycles of Ar$^+$ sputtering (700~eV, 9~mA emission current, $2 \times 10^{-6}$~Torr Ar, 20~min) and annealing at 780~$^\circ$C for 20~min [Fig.~\ref{fig:fig2}(b)]. Note that the annealing temperature is measured at the sample stage; the actual temperature of the crystal is expected to be below 550~$^\circ$C, since no visible thermal radiation from the crystal is observed during annealing. Hereinafter, we refer to the four surface conditions compared in this work as follows: as-prepared, the surface immediately after two sputtering--annealing cycles; after 24~h, the as-prepared surface kept in the analysis chamber at 6~K for 24~h; after sputtering, the surface after an additional sputtering step without subsequent annealing; and after annealing, the sputtered surface after a final annealing step. The four conditions follow this order on the same crystal, and LEED, XPS, and ARPES are measured at each stage.

ARPES and angle-integrated ultraviolet photoemission spectra are measured with a monochromatized He discharge lamp (VUV 5k, Scienta Omicron; He-I$\alpha$, $h\nu = 21.22$~eV) and a hemispherical electron analyzer (R4000, Scienta Omicron) at a sample temperature of 6~K. The pass energy and energy step are 10~eV and 2~meV, respectively, except for the as-prepared spectrum in Fig.~\ref{fig:fig4}(a), which is recorded with an energy step of 20~meV. Constant-energy maps are acquired by rotating the sample about the polar axis in steps of 1$^\circ$ and converted to momentum space using $k_\parallel = 0.5123\sqrt{E_\mathrm{k}}\,\sin\theta$, where $E_\mathrm{k}$ is the kinetic energy in eV and $\theta$ is the emission angle. The measured region covers roughly one quadrant of the surface Brillouin zone; the data in Figs.~\ref{fig:fig5}(b)--\ref{fig:fig5}(e) are obtained by symmetrizing the data with respect to the $C_{4v}$ point group of the Cu(001) surface.

XPS is performed in the same analysis chamber with a non-monochromatic Al K$\alpha$ source (RS 40B1, PREVAC; $h\nu = 1486.6$~eV) at a pass energy of 200~eV. LEED patterns are measured in the preparation chamber at an electron energy of 155~eV at room temperature, immediately after each sputtering or annealing step.

\section{Results}
% ============================================================

To examine the surface order after each treatment, we first compare the LEED patterns measured after sputtering and after annealing [Figs.~\ref{fig:fig2}(c) and \ref{fig:fig2}(d)]. Both patterns show the sharp (1$\times$1) diffraction spots expected for the unreconstructed Cu(001) surface, with the four-fold symmetry of the square surface lattice clearly resolved. The spot positions and the overall symmetry are indistinguishable between the two surfaces. Note that no superstructure spots or streaks appear in either case. On the basis of the LEED patterns alone, therefore, the sputtered and annealed surfaces would both be judged as well-ordered Cu(001).

% ============================================================
%  Fig3
% ============================================================
\begin{figure}[b]
  \includegraphics[width=0.48\textwidth]{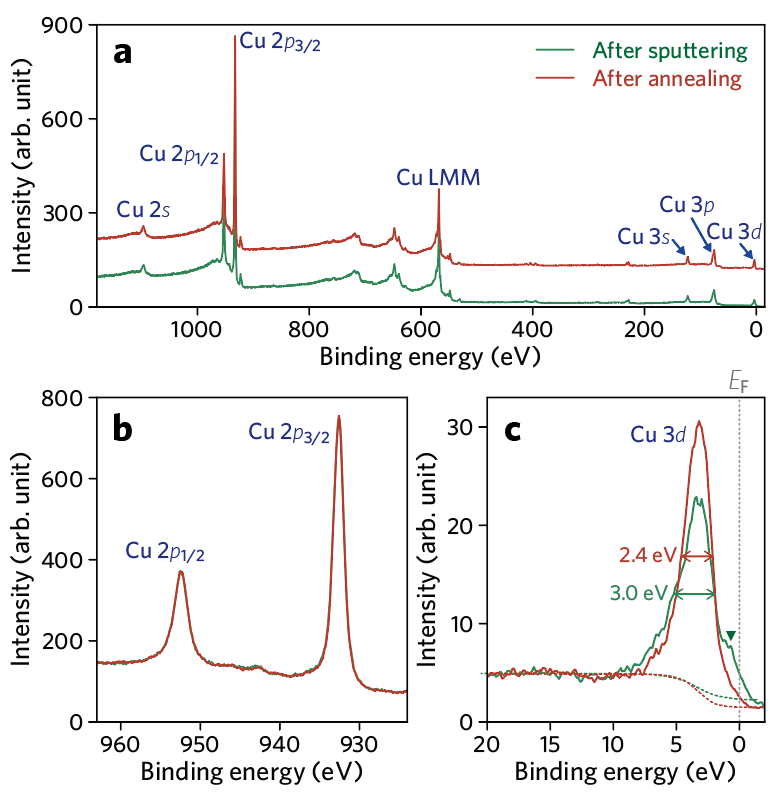}
  \caption{\textbf{XPS after sputtering and after annealing.}
           (a) Survey spectra (non-monochromatic Al K$\alpha$).
           (b) Cu 2$p$ core levels and (c) valence-band region. The core levels are indistinguishable between the two conditions, whereas the Cu 3$d$ valence band is broadened after sputtering and additional intensity marked with a green inverted triangle appears within 2~eV of \EF. Dashed lines in (c) are the Shirley backgrounds, and horizontal bars mark the FWHM of the 3$d$ band for each spectrum. The binding energy is calibrated by setting Cu 2$p_{3/2}$ to 932.6~eV.}
  \label{fig:fig3}
\end{figure}
%===========================================================

% ============================================================
%  Fig4
% ============================================================
\begin{figure*}[t]
\centering
\includegraphics[width=\textwidth]{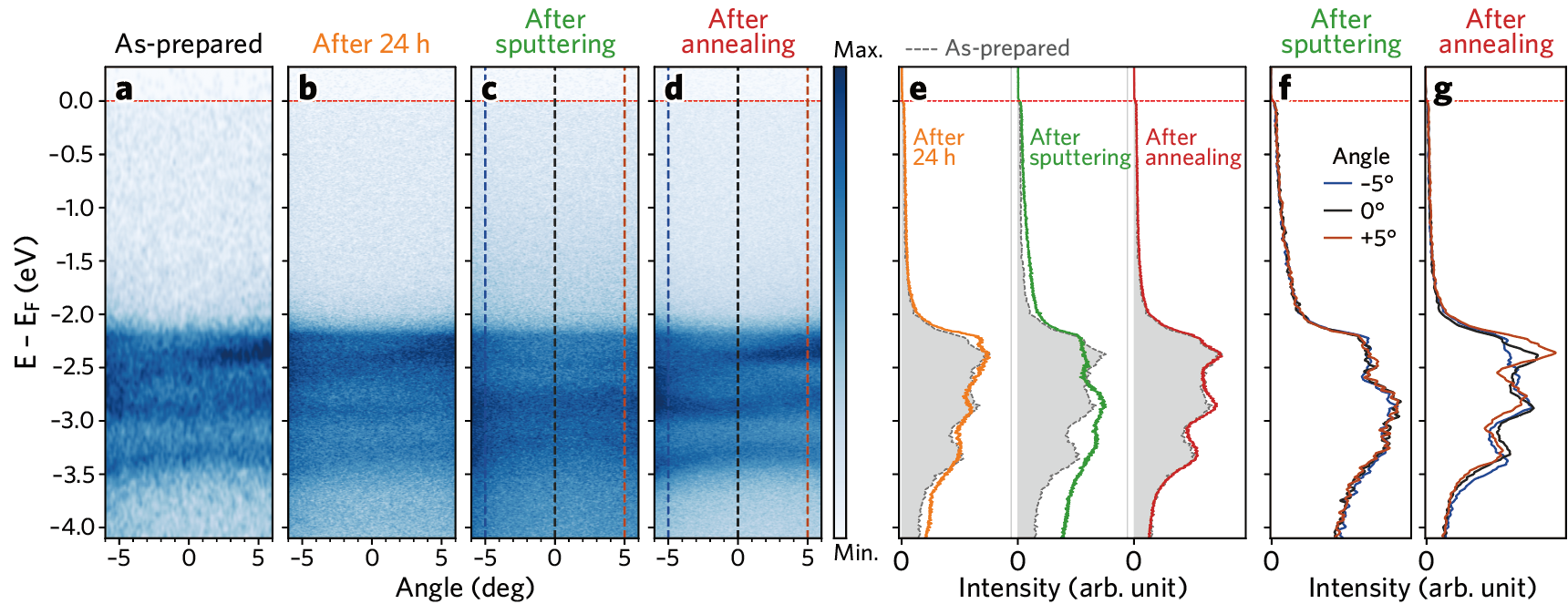}
\caption{\textbf{Angle-resolved photoemission spectra of Cu(001) at each step of the surface treatment.}
(a)--(d) Energy--angle spectra of the as-prepared surface (a), after 24~h in the analysis chamber at 6~K (b), after sputtering (c), and after subsequent annealing (d).
The red dashed line marks $E_\mathrm{F}$, and the blue, black, and brown dashed lines in (c) and (d) mark the emission angles $\theta = -5^\circ$, $0^\circ$, and $+5^\circ$ at which the energy distribution curves (EDCs) in (f) and (g) are taken.
(e) Angle-integrated EDCs ($|\theta| \leq 6^\circ$) after 24~h (orange), after sputtering (green), and after annealing (red), each overlaid on the EDC of the as-prepared surface (gray dashed line with shading); each EDC is normalized to its maximum.
(f), (g) EDCs at $\theta = -5^\circ$ (blue), $0^\circ$ (black), and $+5^\circ$ (brown) after sputtering (f) and after annealing (g).
Each EDC is integrated over $\pm 0.5^\circ$ in $\theta$, smoothed with a Gaussian of $\sigma = 10$~meV, and normalized to its mean intensity between $-4.1$ and $-0.1$~eV. All data are obtained with monochromatized He-I$\alpha$ photons ($h\nu = 21.22$~eV) at 6~K.}
\label{fig:fig4}
\end{figure*}
% ============================================================
Figure~\ref{fig:fig3} shows the XPS spectra measured after sputtering and after annealing. The survey spectra in Fig.~\ref{fig:fig3}(a) exhibit the Cu 2$s$, 2$p$, 3$s$, and 3$p$ core levels, the Cu 3$d$ valence band, and the Cu LMM Auger lines; only negligible C 1$s$ and O 1$s$ signals are detected in either case. Note that the binding energy is calibrated by setting the Cu 2$p_{3/2}$ peak to 932.6~eV~\cite{Chastain1992}, with the same energy offset applied to both spectra, since the Fermi edge is not well resolved at the pass energy used here. The Cu 2$p_{3/2}$ and 2$p_{1/2}$ core levels, enlarged in Fig.~\ref{fig:fig3}(b), overlap almost perfectly between the two surfaces in binding energy, line shape, and intensity. In contrast, the valence-band region in Fig.~\ref{fig:fig3}(c) shows a clear difference. After annealing, the Cu 3$d$ band appears as a single peak with a full width at half maximum (FWHM) of 2.4~eV, estimated after subtraction of a Shirley background~\cite{Shirley1972}. After sputtering, the 3$d$ band broadens to a FWHM of 3.0~eV and loses intensity at its maximum, while additional intensity indicated with a green inverted triangle appears between \EF\ and a binding energy of 2~eV, with a maximum near 0.7~eV.

Figure~\ref{fig:fig4} compares the angle-resolved spectra of the four surface conditions together with the angle-integrated energy distribution curves (EDCs). The as-prepared surface [Fig.~\ref{fig:fig4}(a)] shows the expected Cu(001) valence-band structure: the Cu 3$d$ bands between binding energies of 2 and 4~eV disperse clearly with emission angle. After aging for 24~h in the analysis chamber at 6~K [Fig.~\ref{fig:fig4}(b)], the spectrum remains largely unchanged: the 3$d$ bands remain dispersive, and the low-binding-energy region between \EF\ and 2~eV is nearly identical to that of the as-prepared surface, as confirmed by the orange EDC in Fig.~\ref{fig:fig4}(e). Only a slight increase in intensity around a binding energy of 2.2~eV is discernible.

After sputtering [Fig.~\ref{fig:fig4}(c)], the angle dependence is lost altogether: the intensity between 2 and 4~eV becomes nearly uniform in angle, so that the spectrum becomes momentum-independent and non-dispersive, and additional intensity appears between \EF\ and 2~eV, consistent with the XPS valence band in Fig.~\ref{fig:fig3}(c). After annealing [Fig.~\ref{fig:fig4}(d)], the dispersive 3$d$ bands recover, and the EDC returns to that of the as-prepared surface.

To examine the momentum dependence more directly, we compare in Figs.~\ref{fig:fig4}(f) and \ref{fig:fig4}(g) the EDCs at emission angles of $-5^\circ$, $0^\circ$, and $+5^\circ$ for the sputtered and annealed surfaces. After sputtering [Fig.~\ref{fig:fig4}(f)], the three EDCs coincide within the noise over the entire energy range, including the 3$d$ region; no feature changes its position or intensity with angle. After annealing [Fig.~\ref{fig:fig4}(g)], in contrast, the 3$d$ peaks near binding energies of 2.4, 2.9, and 3.3~eV change in both intensity and position between the three angles, as expected for dispersive bands. Taken together, the four spectra show that prolonged storage at 6~K leaves the angle-resolved spectrum nearly intact, whereas a single sputtering step removes the momentum dependence entirely, and that annealing restores the initial state.

% ============================================================
%  Fig5 (full column)
% ============================================================
\begin{figure*}[t]
  \includegraphics[width=\textwidth]{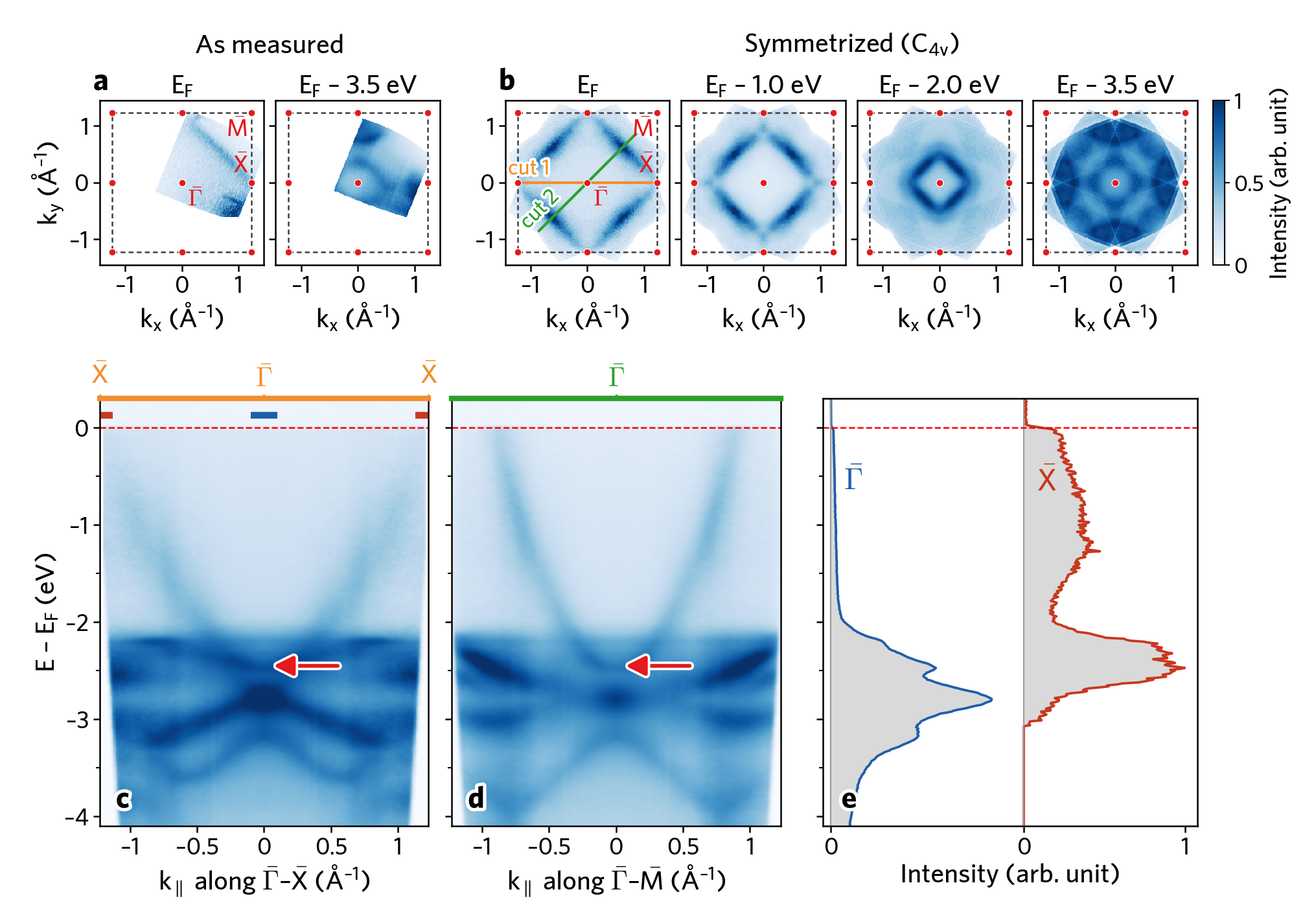}
  \caption{\textbf{ARPES constant-energy maps and high-symmetry cuts of the annealed Cu(001) surface.}
           (a) As-measured maps at \EF\ and at a binding energy of $3.5$~eV.
           (b) $C_{4v}$-symmetrized maps at \EF\ and at binding energies of $1.0$, $2.0$, and $3.5$~eV. All maps are integrated over $\pm 50$~meV. Dashed squares mark the surface Brillouin zone ($\pm\pi/a_s = \pm 1.229$~\invA, where $a_s = a/\sqrt{2} = 2.556$~\AA\ is the surface lattice constant) with \Gb, \Xb, and \Mb\ labeled. Cut~1 (\Gb--\Xb) and cut~2 (\Gb--\Mb) are indicated on the \EF\ map.
           (c) \Gb--\Xb\ and (d) \Gb--\Mb\ cuts ($\pm 0.02$~\invA\ averaged) from the symmetrized data. Red arrows mark the bottom of the $sp$ band at \Gb.
           (e) EDCs integrated within a radius of 0.1~\invA\ around \Gb\ (blue) and \Xb\ (red; averaged over the four equivalent points), each normalized to its maximum. The integration windows are marked above (c). All data are obtained with He-I$\alpha$ at 6~K.}
  \label{fig:fig5}
\end{figure*}
% ============================================================

To characterize the annealed surface in more detail, we map the electronic structure of the annealed Cu(001) surface over the surface Brillouin zone (SBZ) [Fig.~\ref{fig:fig1}(b)]; the resulting three-dimensional data set is shown in Fig.~\ref{fig:fig1}(c), and the constant-energy maps and cuts extracted from it are presented in Fig.~\ref{fig:fig5}. Figure~\ref{fig:fig5}(a) shows the as-measured constant-energy maps at \EF\ and at a binding energy of 3.5~eV, which cover roughly one quadrant of the SBZ. The $C_{4v}$-symmetrized maps at \EF\ and at binding energies of 1.0, 2.0, and 3.5~eV are shown in Fig.~\ref{fig:fig5}(b). At \EF, the map consists of a square-like contour centered at \Gb\ with its corners pointing toward \Xb, together with intense features crossing the \Gb--\Mb\ diagonals. As the binding energy increases to 1.0 and 2.0~eV, the contour shrinks toward \Gb, as expected for the electron-like band whose bottom lies around 2.5~eV [red arrows in Figs.~\ref{fig:fig5}(c) and \ref{fig:fig5}(d)]. At 3.5~eV, the map is dominated by the Cu 3$d$ states, which fill the entire SBZ with a four-fold-symmetric pattern.

Figures~\ref{fig:fig5}(c) and \ref{fig:fig5}(d) show the energy--momentum cuts along \Gb--\Xb\ and \Gb--\Mb\ [cuts 1 and 2 in Fig.~\ref{fig:fig5}(b)]. In both cuts, the dispersive band crossing \EF, i.e., the Cu $sp$ band~\cite{Courths1984}, disperses parabolically from \EF\ down to its bottom at \Gb\ at a binding energy of about 2.5~eV, where it merges into the 3$d$ manifold. The 3$d$ bands occupy binding energies between about 2.2 and 4~eV and show pronounced dispersion along both directions. The EDCs at \Gb\ and \Xb\ in Fig.~\ref{fig:fig5}(e), integrated within a radius of 0.1~\invA\ around each point, summarize this contrast: at \Gb, the spectral weight is concentrated in the 3$d$ region with little intensity near \EF, whereas at \Xb, the $sp$ band contributes a broad feature extending up to \EF\ in addition to the 3$d$ peak. These maps and cuts are consistent with the well-known bulk band structure of Cu projected onto the (001) surface~\cite{Courths1984,Knapp1979,Hufner2003} and confirm that the annealed surface recovers the intrinsic electronic structure of Cu(001).
% ============================================================
\section{Discussion}
% ============================================================

% ============================================================
%  Fig6 (half column)
% ============================================================
\begin{figure}[t]
  \includegraphics[width=0.48\textwidth]{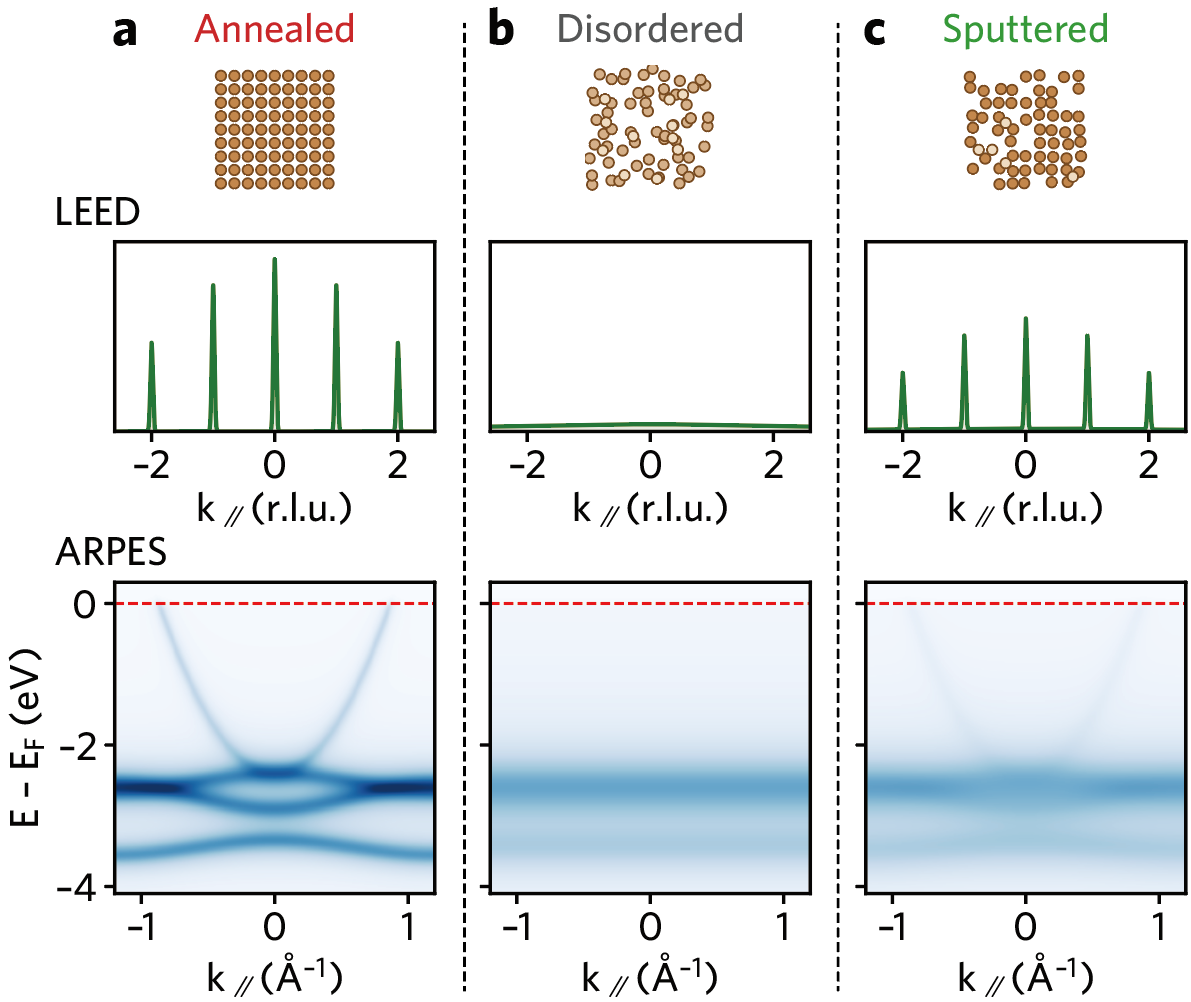}
\caption{\textbf{Schematic illustration of LEED and ARPES on ordered, disordered, and sputtered surfaces.}
(a) Annealed, (b) fully disordered, and (c) sputtered surfaces, the last containing vacancies, adatoms, and their small clusters on the original lattice. From top to bottom: top view of the surface, LEED intensity profile, and ARPES $E$--$k$ cut. The sputtered surface resembles the ordered one in LEED but the disordered one in ARPES. All profiles and intensity maps are qualitative schematics, not calculated spectra.}
  \label{fig:fig6}
\end{figure}
% ============================================================

The central finding of this work is that the sputtered and annealed Cu(001) surfaces are nearly indistinguishable in LEED and in XPS core levels, yet differ qualitatively in ARPES. We first consider why the three probes respond so differently to the same surface.

LEED and XPS core-level spectroscopy can be both insensitive to the type of disorder introduced by sputtering, but for different reasons. In LEED, the diffraction spots arise from the average periodic lattice, whereas vacancies, adatoms, and their small clusters contribute to a diffuse background spread over the whole screen~\cite{Henzler1982, Broekmann2002}. Because the sputtered surface is still attached to the single-crystal substrate and most Cu atoms remain on the lattice sites of the underlying crystal, the spots remain sharp and are expected to lose part of their intensity to the background [Fig.~\ref{fig:fig6}(c)]. The XPS core-level spectra, in contrast, reflect the chemical state and composition of the near-surface region: Ar$^+$ sputtering removes material but does not tend to change the chemical identity of the Cu atoms, so the Cu 2$p$ lines remain unchanged, as observed in Fig.~\ref{fig:fig3}(b)~\cite{Biesinger2017}. In short, XPS core levels are sensitive to the composition but insensitive to the structural order of the surface.

Unlike LEED, which probes the atomic positions, ARPES probes the momentum-resolved electronic states, and therefore responds to the same defects in a different way. On a well-ordered surface, the parallel momentum $k_\parallel$ of the photoelectron is conserved, so that the spectrum at each emission angle reflects the bands at the corresponding $k_\parallel$ [Fig.~\ref{fig:fig6}(a)]. Vacancies, adatoms, and their small clusters scatter the electrons, which broadens the bands in momentum and transfers spectral weight to a momentum-independent contribution~\cite{Shevchik1977}, as on a fully disordered surface [Fig.~\ref{fig:fig6}(b)]. Consequently, a sputtered surface can resemble an ordered surface in LEED and a disordered one in ARPES [Fig.~\ref{fig:fig6}(c)].

Our data are consistent with this picture. The EDCs of the sputtered surface at different emission angles coincide within the noise [Fig.~\ref{fig:fig4}(f)], whereas those of the annealed surface differ clearly [Fig.~\ref{fig:fig4}(g)]. The valence-band XPS supports this interpretation [Fig.~\ref{fig:fig3}(c)]: after sputtering, the Cu 3$d$ band broadens and develops intensity within 2~eV of $E_\mathrm{F}$, as expected for Cu atoms in a modified local environment~\cite{Hammer1995,Ruban1997}, and the same intensity appears in the He-I$\alpha$ spectrum of the sputtered surface [Fig.~\ref{fig:fig4}(c)] but not after annealing.

Taken together, LEED spots survive sputtering because the average lattice remains periodic, and XPS core levels survive because they tend to depend on composition, whereas the band dispersion in ARPES is lost because the electrons are scattered by the defects in the topmost layers. The three probes thus report on three different aspects of the same surface.

From a practical point of view, our results indicate that a sharp LEED pattern and clean core-level spectra are necessary but not sufficient conditions for a surface suitable for ARPES, and that the quality of the final annealing step must be judged from the angle-resolved spectrum itself. We also note that keeping the as-prepared surface at 6~K for a day does not noticeably degrade the angle-resolved spectrum, whereas a single sputtering step removes the dispersion entirely. Further studies may be desired to quantify the relation between the LEED spot profile and the loss of dispersion of the bulk bands in a more systematic manner, as has been done for the linewidth of surface states~\cite{Theilmann1997}.

% ============================================================
\section{Conclusion}
% ============================================================

In summary, we compare LEED, XPS, and ARPES of a Cu(001) single crystal measured \textit{in situ} at four stages of surface preparation: as-prepared, after aging for 24~h at 6~K, after sputtering, and after annealing. The sputtered and annealed surfaces are nearly indistinguishable in LEED and in the Cu 2$p$ core levels, whereas in ARPES the sputtered surface loses the band dispersion entirely and the annealed surface recovers the full Cu(001) band structure. We attribute this contrast to how each probe responds to sputtering-induced defects: LEED reflects the average lattice, which remains periodic, XPS core levels reflect the composition, which is unchanged, and ARPES reflects the electronic states, whose crystal momentum is no longer well defined. Our results show that a sharp LEED pattern and clean core-level spectra do not by themselves guarantee a surface suitable for ARPES, and that the ARPES spectrum is more sensitive indicator of the surface quality relevant to band-structure measurements.

% ============================================================
\begin{acknowledgments}
We gratefully acknowledge insightful discussions with Dr. Younsik Kim, Dr. Sae Hee Ryu, and Dr. Jongkeun Jung. This work was supported by the National Research Foundation of Korea (NRF) grant funded by the Korea government (MSIT) (Grant No. RS-2025-00513951) and by the Korea Basic Science Institute (KBSI) Young Researcher Infrastructure Support Program funded by the MSIT (Grant No. RS-2025-00612977). The oxide MBE system connected to the analysis chamber used in this work was supported through the National Research Facilities \& Equipment Center (NFEC) (No. NFEC-2026-03-313997).
\end{acknowledgments}

\bibliographystyle{apsrev4-2}
\bibliography{ref}

\end{document}